\documentstyle[emulateapj,psfig]{article}

\begin{document}
\title{Temporal Profiles and Spectral Lags of XRF 060218}
\author{
En-Wei Liang$^{1,2}$, Bin-Bin Zhang $^{1,3,4}$, Mike Stamatikos$^{5}$,
Bing Zhang$^1$, Jay Norris$^5$, Neil Gehrels$^5$, Jin Zhang $^{3,4}$,
Z. G. Dai$^{6,1}$}

\affil{$^1$Department of Physics, University of Nevada, Las Vegas, NV
89154, USA; \\lew@physics.unlv.edu;bzhang@physics.unlv.edu\\
$^2$Department of Physics, Guangxi University, Nanning 530004, China\\
$^3$National Astronomical Observatories/Yunnan Observatory, CAS,
Kunming 650011, China\\
$^4$The Graduate School of the Chinese Academy of Sciences,
Beijing 100039, China\\
$^5$NASA Goddard Space Flight Center, Greenbelt, MD 20771\\
$^6$Department of Astronomy, Nanjing University, Nanjing 210093, China}

\begin{abstract}
The spectral and temporal properties of the non-thermal emission of
the nearby XRF 060218 in 0.3-150 keV band are studied. We show that
both the spectral energy distribution and the light curve properties
suggest the same origin of the non-thermal emission detected by {\em
Swift} BAT and XRT. This event has the longest pulse duration
and spectral lag observed to date among the known GRBs. The pulse
structure and its energy dependence are analogous to typical GRBs. By
extrapolating the observed spectral lag to the {\em CGRO/BATSE} bands
we find that the hypothesis that this event complies with the same
luminosity-lag relation with bright GRBs cannot be ruled out at
$2\sigma$ significance level. These intriguing facts, along with its
compliance with the Amati-relation, indicate that XRF 060218 shares
the similar radiation physics as typical GRBs.
\end{abstract}

\keywords{gamma-rays: bursts---method: statistical}


\section{Introduction}
X-ray flashes (XRFs), cosmic explosions with lower spectral peak energies ($E_p$) in $\nu
F_\nu$ spectra than typical gamma-ray bursts (GRBs) (Heise et al. 2001; Kippen et al.
2003), are thought to be the low energy extension of typical GRBs (Lamb et al. 2005;
Sakamoto et al. 2004, 2006; Cui et al. 2005). They may be GRB jets (uniform or
structured) viewed at off-axis directions (e.g. Zhang et al. 2003; Yamazaki et al. 2004;
Zhang et al. 2004) or intrinsically different events (Lamb et al. 2005; Soderberg et al.
2005). It is long speculated that long duration GRBs originate from a relativistic jet
emerging from a collapsing massive star progenitor (Woosley et al.  1993; Paczynski 1998;
MacFadyen et al. 1999; Zhang et al. 2003), and associations of core-collapsing supernovae
(SNe) with long GRB afterglows have been spectroscopically identified in a number of
systems, including GRB980425/SN1998bw (Galama et al. 1998), GRB030329/SN2003dh (Stanek et
al.  2003; Hjorth et al. 2003), and GRB031203/SN 2003lw (Malesani et al. 2004). The firm
spectroscopic association of the nearby XRF 060218 (at z=0.0331, Mirabal et al. 2006)
detected by Swift (Campana et al. 2006) with the Type Ic SN 2006aj was established
recently (Modjaz et al.  2006; Pian et al. 2006; Sollerman et al. 2006; Mirabal et al.
2006; Cobb et al. 2006; Soderberg et al. 2006). This suggests that the progenitors of
both GRBs and XRFs are related to the death of massive stars and XRFs are the low energy
extension of the more ``standard'' GRBs.

Some empirical relations have been discovered from typical GRBs. It is interesting to
verify whether XRF 060218 satisfies these relations as an approach to access its
``standardness''. It has been reported that XRF 060218 complies with the isotropic energy
vs. spectral peak energy ($E_{iso}-E_p$) relation derived from long GRBs (Amati et al.
2002) and other XRFs (Amati et al. 2006). The pulses in GRB light curves usually display
a fast-rise-exponential-decay (FRED) shape and it has been found that the pulse width is
related to the energy of the observational band as $\omega \propto E^{-0.4}$ (Fenimore et
al. 1995; Norris et al. 2005). A correlation between the isotropic luminosity and the
spectral lag ($L_{\rm iso}-\tau$ relation) of light curves was also discovered with
typical GRBs as $L_{\rm iso} \propto \tau^{-1.18}$ (Norris et al. 2000). Both $\omega-E$
and $L_{\rm iso}-\tau$ relations are related to the structure of light curves. Since most
photons of an XRF are in the X-ray band, previous GRB missions did not observe the real
light curves of XRFs. This makes it difficult to identify the temporal structure and the
spectral lag of XRFs. XRF 060218 is long and soft, and {\em Swift} BAT and XRT
simultaneously collected the data in the gamma-ray to X-ray bands. This makes it possible
to measure its temporal structure and to examine whether it complies with the same
$\omega-E$ and $L_{\rm iso}-\tau$ relations derived from GRBs. In this {\em Letter} we
focus on this issue. Throughout the paper $H_0=71$ km s$^{-1}$ Mpc$^{-1}$,
$\Omega_{\rm{m}}=0.3$, and $\Omega_\Lambda=0.7$ are adopted.

\section{Data}
XRF 060218 was detected with the {\em Swift}/BAT on 2006 February
18.149 UT. It is a long burst, with a duration $T_{90}\sim 2000$
seconds in the 15-150 keV band. Swift slewed autonomously to the burst
and the X-ray telescope (XRT) and UV/Optical Telescope (UVOT) began
collecting data 159 s after the burst trigger.

\begin{figure}
\plotone{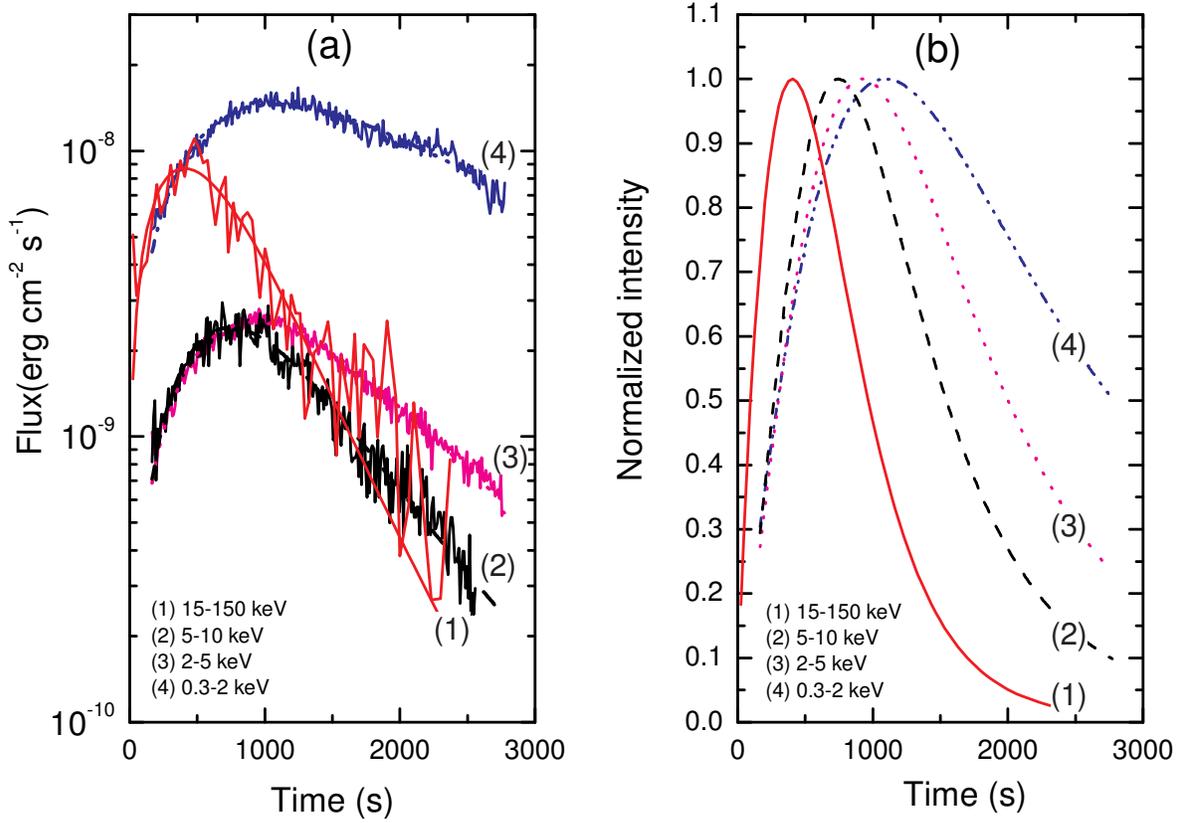}
\caption{(a) Unabsorbed light curves of the non-thermal
gamma-rays/X-rays in the energy bands of 15-150 keV, 5-10 keV, 2-5
keV, and 0.3-2 keV, respectively. The fitting curves with Eq.(1) are
plotted. (b) Normalized light curves from the empirical model fitting.
\label{LC}}
\end{figure}

XRF 060218 was an image trigger. The BAT event data lasted only until $t\sim 300$ seconds
after the trigger. The survey data are used to derive the BAT light curve. XRF 060218 is
very soft, with most of the emission in the BAT band being lower than 50 keV (Campana et
al. 2006). In order to obtain a high level of signal-to-noise ratio of the BAT light
curve, we use the light curve in whole BAT band, i.e., 15-150 keV. The first orbit of the
XRT data is fully in the Windowed Timing (WT) mode. We extract the light curves and
spectrum of the XRT data with the Xselect package. The spectrum is grouped with the tool
{\em grppha}, and the spectral fitting is carried out with the {\em Xspec} package.
Following Campana et al. (2006) we fit the XRT spectrum in the first orbit with a model
combining a black body component with temperature $kT$ and a cut-off power law ($F\propto
E^{-\Gamma}e^{-E/E_{c}}$) component. The absorption in both Milky Way\footnote{The $N_H$
of our Galaxy (Dickey et al. 1990) for this burst is $\sim 1.1\times 10^{21}$ cm$^{-2}$}
and the GRB host galaxy ($N_H^{host}$) is incorporated.  We obtain
$N_H^{host}=0.63^{+0.03}_{-0.03}\times 10^{22}$ cm$^{-2}$, $kT=0.122^{+0.003}_{-0.004}$
keV, $\Gamma=1.78^{+0.08}_{-0.08}$, and $E_{c}=21.8^{+15.8}_{-6.7}$ keV (corresponding to
$E_p\sim 5$ keV), with reduced $\chi^2=1.55$ for 769 degrees of freedom. We derive the
unabsorbed light curves in the 0.3-2 keV, 2-5 keV, and 5-10 keV bands from the XRT data.
As reported by Campana et al.  (2006), a thermal component exists, and is likely of
different origin (e.g. the shock breakout emission associated with SN 2006aj; c.f. Li
2006; Ghisellini et al. 2006b) from the non-thermal component. We therefore subtract the
contribution of this component from the observed light curves. The $kT$ and the radiation
radii ($R_{BB}$) evolve with time during the first orbit. We read the values of $kT$ and
$R_{BB}$ from Campana et al.(2006) and calculate the light curves of this component in
0.3-2, 2-5, and 5-10 keV bands. Since the temperature of the black body component is
below $0.2$ keV, the light curves in the energy band higher than 2 keV is essentially not
contaminated by the thermal component. In the 0.3-2 keV band, the derived light curve of
the thermal component continuously increase with time, which could be well fitted by
$\log F_{0.3-2}=(-10.69\pm 0.09)+(0.66\pm 0.03)\log t$ in cgs units. We thus subtract the
contribution of this component from the 0.3-2 keV band light curve according to this
fitting result. The derived non-thermal light curves in the three XRT bands as well as
the one in the BAT band are shown in Fig.\ref{LC}(a). To clearly view the pulse width and
spectral lag dependences with energy, we also plot in Fig.\ref{LC}(b) the normalized
lightcurves. The characteristics of the light curves along with the average photon energy
($\bar{E}$) of each energy band are reported in Table 1.

\section{Results}
\subsection{Spectral energy distribution}
\begin{figure}
\plotone{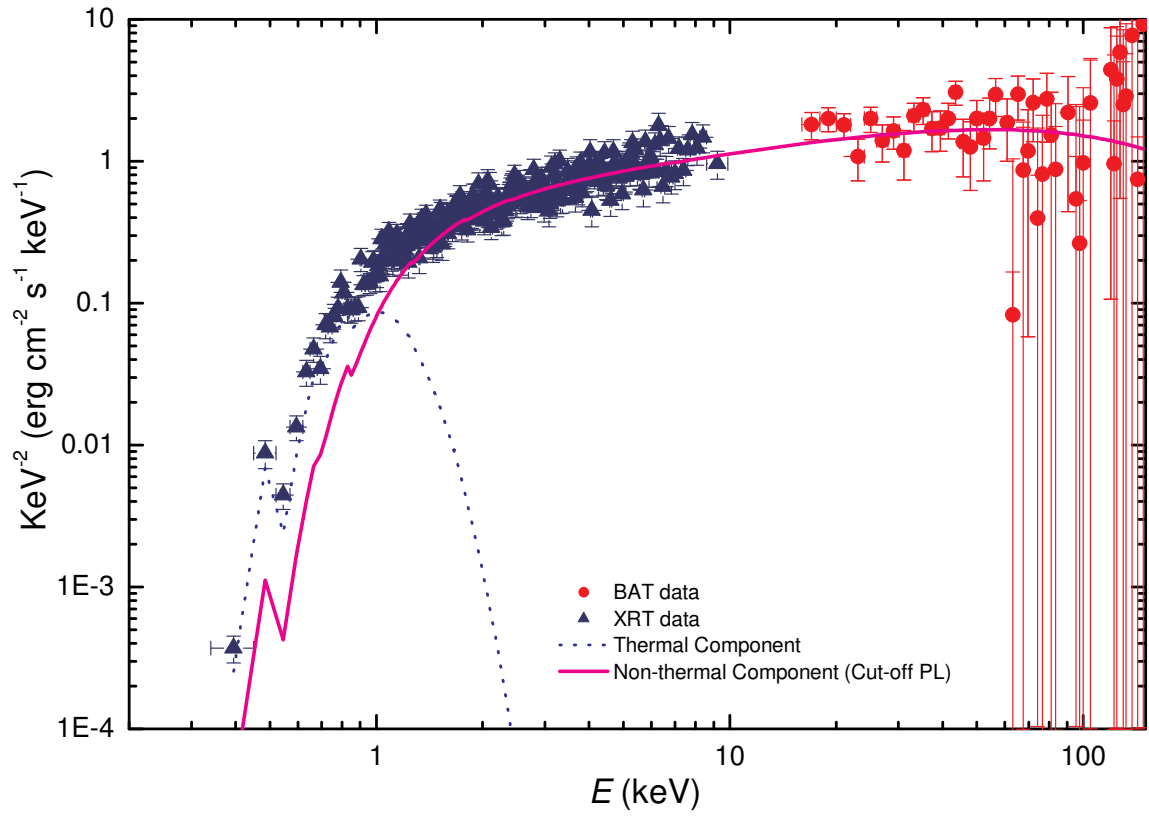}
\caption{The BAT-XRT joint spectral energy distribution from 160 to
300 seconds since the BAT trigger. The BB+CutoffPL fitting model is
also shown.}
\label{SED}
\end{figure}

One important question is whether the XRT non-thermal emission and the
BAT emission are of the same origin. In order to clarify this we first
study the spectral energy distribution (SED) of the burst using joint
BAT-XRT data. Since this event is image trigger, the BAT event file
contains only the data in the first 300 seconds since the BAT
trigger. XRT began collecting data 159 seconds after the BAT
trigger. We thus only obtain simultaneous observations of the two
instruments from 160 to 300 seconds with the event files. The
joint-fit SED in this period is shown in Fig.\ref{SED}, which is well
fitted by the BB+CPL model (Campana et al. 2006) with the following
parameters: $N_H^{host}=0.65^{+0.03}_{-0.03}\times 10^{22}$ cm$^{-2}$,
$kT=0.13^{+0.01}_{-0.01}$ keV, $\Gamma=1.56^{+0.08}_{-0.07}$ and
$E_{c}=122^{+160}_{-46}$ keV (corresponding to $E_p\sim 54$ keV), with
reduced $\chi^2=0.97$ for 327 degrees of freedom. The $E_p$ strongly
evolves with time, from $54$ keV at the beginning down to $\sim
5$ keV at later times. This is consistent with that reported by Campana
et al. (2006) and Ghisellini et al. (2006a). The model
fitting results are shown in Fig.\ref{SED}. One can observe that the
BAT-component is a good extrapolation of the XRT non-thermal
component.  This result implies that the non-thermal emissions
detected by XRT and BAT are of the same origin.

\subsection{Pulse width and Energy Dependence}
\begin{figure}
\epsscale{1.0}
\plotone{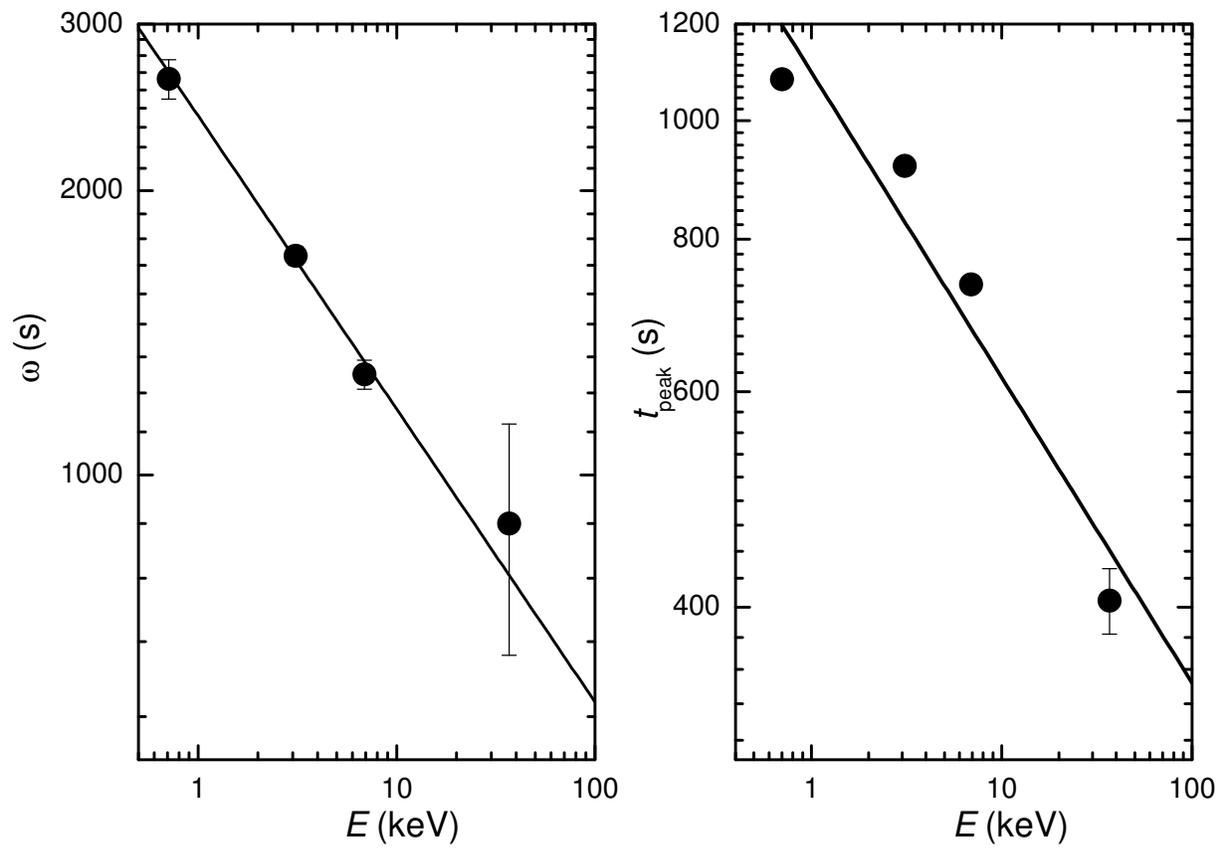}
\caption{The pulse duration (panel a) and the peak
time (panel b) as a function of the average photon energy of the
non-thermal emission. The solid lines in both panels are the best
fits.}
\label{FWHM_LAG}
\end{figure}

As shown in Fig.\ref{LC}, the light curves in different energy bands
can be all modeled by a single FRED-pulse. Kocevski et al.  (2003)
developed an empirical expression to fit a FRED-like pulse, which
reads,
\begin{equation}\label{Kocevski}
F(t)={F_m}(\frac{t+t_0}{t_m+t_0})^r[\frac{d}{d+r}+\frac{r}{d+r}(\frac{t+t_0}
{t_m+t_0})^{(r+1)}]^{-\frac{r+d}{r+1}},
\end{equation}
where $t_{m}$ is the time of the maximum flux ($F_{m}$), $t_0$ is the
offset time, and $r$ and $d$ are the rising and decaying power-law
indices, respectively. We fit the light curves with
Eq.(\ref{Kocevski}) and then measure the pulse width, rising and
decaying times at the full-width half-maximum (FWHM) of the fitting
light curves, and the rising-to-decaying time ratio ($\varphi$). The
errors of these quantities are derived from simulations by assuming a
normal distribution of the errors of the fitting parameters.  The
reported errors are at $1\sigma$ confidence level. The results are
tabulated in Table 1. We show $\omega$ as a function of $\bar{E}$ in
Fig.\ref{FWHM_LAG}(a). Apparently the two quantities are correlated. A
best fit yields $\omega\propto E^{-0.31\pm 0.03}$.  The $\varphi$
parameter ranges from $0.43$ to 0.59. It is found that XRF 060218 roughly
satisfies the same $\omega - E$ relation (Fenimore et al. 1995; Norris
et al. 2005) and its $\varphi$ values are also well consistent with
that observed in typical GRBs (e.g. Norris et al. 1996; Liang et
al. 2002), although it has a much longer pulse width than other
single-pulse GRBs. These results imply that XRF 060218 may be an
extension of GRBs to the extremely long and soft regime.

\subsection{Spectral Lag and Energy Dependence}
The light curves shown in Fig.\ref{LC}(a) display a significant
spectral lag ($\tau$), with soft photons lagging behind the hard
photons, as usually seen in long GRBs (Norris et al. 2000; Yi et
al. 2006). We illustrate this lag behavior with the
intensity-normalized light curves in Fig.\ref{LC}(b). The light curves
peak at $405\pm 25$, $735\pm 9$, $919\pm 7$, and $1082\pm 13$ seconds,
respectively, in a sequence of high energy band to low energy band as
shown in Fig. \ref{FWHM_LAG}(b). The best fit to the correlation
between the peak time ($t_{peak} $) and the average photon energy
yields,
\begin{equation}\label{T_E}
\log t_{peak}=(3.04\pm 0.04)-(0.25\pm0.05) \log E.
\end{equation}
A simple estimate of the lags between any pairs of the 4 light curves obtains
$\tau_{peak}=163\sim 677$ seconds, being consistent with that shown in Gehrels et al.
(2007). We note XRF 060218 becomes the new record-holder of the long-lag, wide-pulse
GRBs. The previous record holder was GRB 971208, with $\tau\sim 58$ seconds and
$\omega=395$ seconds (Norris et al. 2005).

\begin{figure}
\plotone{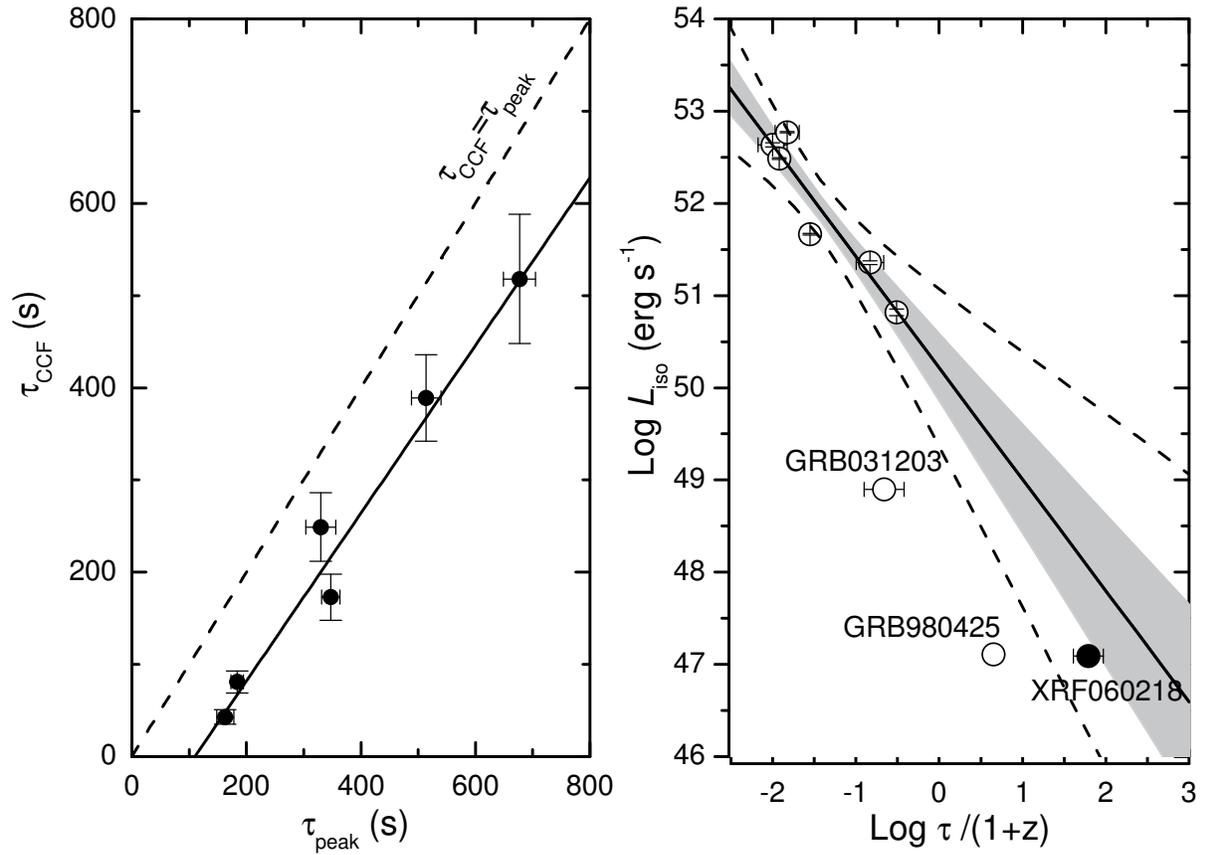}
\caption{Panel (a): Comparison of the spectral lags derived from the
peak times and from the CCF method. The solid line is the best
fit. Panel (b): Isotropic gamma-ray luminosity as a function of
spectral lag. The spectral lags of typical GRBs and GRB 980425 are
calculated with the light curves in the 25-50 keV and 100-300 keV
bands observed by CGRO/BATSE. The lag of GRB 031203 is calculated with
the light curves in the 20-50 keV and 100-200 keV bands. The grey band
and the two dashed lines mark the best fits at the $1\sigma$ and $2\sigma$
confidence level, respectively, and the solid line is the regression
line for the six typical GRBs presented in Norris et al. (2000).}
\label{Liso_lag}
\end{figure}

We also calculate the lags with the cross correlation function (CCF) method. The errors
of lags are evaluated by simulations. The results are also reported in Table 1. The lag
derived by the CCF method ($\tau_{\rm CCF}$) is strongly correlated with $\tau_{\rm
peak}$, but is systematically lower\footnote{We here derive $\tau_{\rm CCF}$ from the
peak of the CCF without considering the side lobe contribution of the CCF.  A fit to the
CCF with a cube or quartic function gives a larger lag by considering the side lobe
contribution (Norris et al. 2000). However, this method strongly depends on the
artificially-selected range of CCF for the fitting. Since the light curves are a smooth
pulse and their lags are significantly larger than the time bin, the peaks of CCFs are
robust to estimate the lags.} than $\tau_{\rm peak}$ [Fig. \ref{Liso_lag}(a)]. A best fit
gives $\tau_{\rm CCF}=(-100\pm 17)+(0.91\pm 0.08)\tau_{\rm peak}$.

The $L_{\rm iso}-\tau$ relation was discovered with six bright BATSE
GRBs (Norris et al. 2000) and the spectral lag was defined by the
light curves in the 25-50 keV and 100-300 keV bands. We investigate
whether the lag behavior of XRF 060218 is consistent with the $L_{\rm
iso}-\tau$ relation. Since XRF 060218 is a soft XRF and the emission
in the 100-300 keV band is too weak to derive a light curve,
we assume that $t_{peak}$ of the light curve in the 100-300
keV band follows the $t_{peak}-E$ relation (Eq.\ref{T_E}) and
perform the extrapolation.  With the extrapolated $t_{peak}$ we
then estimate $\tau_{peak}$ for the light curves in the 25-50 keV
(average energy 30 keV) and 100-300 keV (average energy 200 keV)
bands. We obtain $\tau_{peak}=177\pm 16$ seconds. Since
$\tau_{\rm CCF}$ is more reliable, we use the $\tau_{\rm peak}
- \tau_{\rm CCF}$ relation [Fig. \ref{Liso_lag}(a)] to derive
$\tau_{\rm CCF}=61\pm 26$ seconds. This lag is used in the $L-\tau$
relation analysis. Using the peak fluxes in the BAT and XRT band,
we estimate $L_{iso}=1.2\times 10^{47}$ ergs s$^{-1}$.

Figure \ref{Liso_lag}(b) shows the $L_{iso}-\tau$ relation derived by Norris et al.
(2000) compared against XRF 060218 as well as two other nearby GRBs, 980425 and 031203.
The data of the previous GRBs are taken from Norris et al. (2000) and Sazonov et al.
(2004)\footnote{The peak photon fluxed in the 50-300 keV bands of these GRBs are
converted to energy fluxes with the spectral index presented in Friedman \& Bloom (2005),
and their $L_{iso}$'s are recalculated with the cosmological parameters used in this
paper.}. The grey band and the two dashed lines mark the best fits at the $1\sigma$ and
$2\sigma$ confidence level, respectively, and the solid line is the regression line for
the 6 GRBs that were used to draw the $L-\tau$ correlation, i.e. $\log L_{\rm
iso}=(50.22\pm 0.32)-(1.21\pm 0.21)\times \log \tau$ (errors are at the $1\sigma$ level).
We can see that XRF 060218 is definitely inside the $2\sigma$ region and is marginally at
the $1\sigma$ region boundary. Therefore, the hypothesis that XRF 060218 follows the
$L-\tau$ relation cannot be ruled out at the $2\sigma$ significance level. We caution
that the $\tau$ is inferred from the extrapolation of the $t_{\rm peak}-E$ relation. This
introduces uncertainties in deriving the lag. The other two nearby GRBs, 980425 and
031203, are out of the $2 \sigma$ region, which are identified as significant outliers of
this relation (e.g. Sazonov et al. 2004).

\section{Conclusions and discussion}

We have investigated the non-thermal emission of XRF 060218. The early SED of this event
from 0.3-150 keV observed by BAT and XRT suggests that the non-thermal emission detected
by the two instruments are the same component. By subtracting the contribution of the
thermal emission we derive the light curves of the non-thermal emission. They are
composed of a broad single pulse, and the energy dependences of the widths and the
rising-to-decaying-time ratio of the pulses are roughly consistent with those derived in
typical GRBs. The light curves show significant spectral lags, with a well-defined peak
time sequence from high energy band to low energy bands, i.e. $t_{\rm peak}\propto
E^{-0.25\pm0.05}$. We infer the spectral lag in the BATSE bands and find that the
hypothesis that this event complies with the $L_{\rm iso}-\tau$ relation with typical
GRBs cannot be ruled out at the $2\sigma$ significance level.

These intriguing facts, along with its compliance with the Amati-relation, strongly
suggest that GRB 060218 is a ``standard'' burst at the very faint, long, and soft end of
the GRB distribution. Since all these relations concern the temporal and spectral
properties of emission, they are likely related to the radiation mechanisms. The results
therefore imply that XRF 060218 and other XRFs may share the similar radiation physics
(e.g. synchrotron or inverse Compton scattering in internal shocks, M\'{e}sz\'{a}ros,
2002; Zhang \& M\'{e}sz\'{a}ros 2004; Piran 2005; M\'{e}sz\'{a}ros 2006) with harder GRBs
.

As discovered by Norris (2002), the proportion of long-lag bursts within long-duration
bursts increases from negligible among bright BATSE bursts to $\sim 50\%$ at the trigger
threshold, and their peak fluxes are $\sim 2$ orders of magnitude lower than those of the
brightest bursts. This argues that they are intrinsically under-luminous. Taken together
with the fact that three nearby GRBs, 980425, 031203, and 060218, are long-lagged and
under-luminous, an intuitive speculation is that long-lag bursts are probably relatively
nearby (e.g., Norris et al. 2005). The local GRB rate of these GRBs thus should be much
higher than that expected from the high luminosity GRBs (Liang et al. 2006; see also Cobb
et al. 2006; Pian et al. 2006; Soderberg et al. 2006). A possible scenario to explain
their wide-pulse, long-lag, and under-luminous features is the off-axis viewing angle
effect (e.g. Nakamura 1999; Salmonson 2000; Ioka \& Nakamura 2001).  Another scenario is
that these features are intrinsic, being due to their lower Lorentz factors (Kulkarni et
al. 1998; Woosley \& MacFadyen 1999; Salmonson 2000; Dai et al. 2006; Wang et al. 2006).
They might be from a unique GRB population (Liang et al. 2006) having a different type of
central engine (e.g. neutron stars rather than black holes) from bright GRBs (e.g.
Mazzali et al. 2006; Soderberg et al. 2006).

\acknowledgments
We thank the anonymous referee for helpful
suggestions, and S. Campana, D. Burrows, J. Nousek, K. Page,
T. Sakamoto, X.-Y. Wang, and Z. Li for discussion. This work was
supported by NASA under grants NNG06GH62G and NNG05GB67G, and the
National Natural Science Foundation of China under grants
10463001(EWL).

\clearpage

\begin{table}
\caption{Temporal structures and  spectral lags  of the light curves} Temporal Structures

\begin{tabular}{lllllll}
\hline
Band & Peak(s) &$\omega$ (s) & Rising time (s)& Decaying time (s) & $\varphi$&$\bar{E}$ (keV)\\
\hline
(1)15-150 keV  &405(25)   &  889 (244)     &  311 (28)  &    578(185)  & 0.54(0.18)  &    36.9 \\
(2)5-10 keV    &735(9)    &  1278 (45)    &  475 (12)  &    803(35)   & 0.59(0.03)   &    6.9\\
(3)2-5 keV     &919(7)    &  1707 (40)    &  624 (8)   &    1084(34)  & 0.58(0.02)   &   3.1\\
(4)0.3-2 keV   &1082(13)  &  2625 (125)   &  794 (14)  &    1831 (112)& 0.43(0.03) &    0.7\\
\hline
\end{tabular}
\\
\hspace{6pt}

Spectral Lags

\begin{tabular}{llllllll}
\hline
Bands & $\Delta E$(keV)&$\tau_{\rm peak}$(s)&$\tau_{\rm CCF}$(s)&Bands & $\Delta E$(keV) &$\tau_{\rm peak}$(s)&$\tau_{\rm CCF}$(s)\\
\hline
(1)-(2)& 30.0  &330(26)&249(37)  &(1)-(3)&33.8 &514(26)& 389(47) \\
(1)-(4)&36.2  &677(28)&518(70) &(2)-(3)& 3.8 &184(11)&81(12)\\
(2)-(4)& 6.2 &347(16)&173(25)  &(3)-(4)&2.4  &163(15)&43(8) \\
\hline
\end{tabular}

Note: all the errors are derived by simulations and in $1\sigma$ significance level.
\end{table}

\clearpage

\clearpage

\clearpage

\clearpage

\end{document}